\documentclass[10pt,conference]{IEEEtran}
\usepackage{epsfig,setspace,amsmath,epsf,amssymb,bm,theorem,cite,graphicx, epstopdf,algorithm,algpseudocode,float,color,mathtools}
\usepackage[table,xcdraw]{xcolor}

\newtheorem{theorem}{Theorem}

\newtheorem{remark}{Remark}

\newcommand{\diag}{\mathop{\mathrm{diag}}}

\IEEEoverridecommandlockouts
\allowdisplaybreaks

\begin{document}

\title{Rate-Privacy-Storage Tradeoff in Federated Learning with Top $r$ Sparsification}

\author{Sajani Vithana \qquad Sennur Ulukus\\
	\normalsize Department of Electrical and Computer Engineering\\
	\normalsize University of Maryland, College Park, MD 20742 \\
	\normalsize \emph{spallego@umd.edu} \qquad \emph{ulukus@umd.edu}}

\maketitle

\begin{abstract}
We investigate the trade-off between rate, privacy and storage in federated learning (FL) with top $r$ sparsification, where the users and the servers in the FL system only share the most significant $r$ and $r'$ fractions, respectively, of updates and parameters in the FL process, to reduce the communication cost. We present schemes that guarantee information theoretic privacy of the values and indices of the sparse updates sent by the users at the expense of a larger storage cost. To this end, we generalize the scheme to reduce the storage cost by allowing a certain amount of information leakage. Thus, we provide the general trade-off between the communication cost, storage cost, and information leakage in private FL with top $r$ sparsification, along the lines of two proposed schemes. 
\end{abstract}

\section{Introduction}
In federated learning (FL) \cite{FL1,FL2}, a set of users remotely train an ML model using their own local data in their own devices, and share only the gradient updates with the central server. This reduces the privacy leakage of users while decentralizing the processing power requirements of the central server. However, it has been shown that the gradients shared by a user leak information about the user's private data \cite{comprehensive,MembershipInterference,SecretSharer,featureLeakage,InvertingGradients,DeepLeakage}. 

Apart from the privacy leakage, another drawback of FL is the significantly large communication cost incurred by sharing model parameters and updates with millions of users in multiple rounds. One solution to this problem is gradient sparsification \cite{sparse1,adaptive,rtopk}, where the users only communicate a selected set of gradients and parameters as opposed to communicating all gradient updates and parameters. Top $r$ sparsification \cite{rtopk} is a widely used sparsification technique, where only the most significant $r$ fraction of parameters/updates are shared between the users and the central server, which significantly reduces the communication cost, since $r$ is typically around $10^{-2}$ to $10^{-3}$.  

However, in sparsified FL, the values as well as the positions of the sparse updates leak information about the user's local data. Note that the positions of the sparse updates convey information about the most and least significant sets of parameters for a given user leaking information about its private data. Thus, in order to guarantee the privacy of users participating in the sparse FL process, two components need to be kept private, namely, 1) values of sparse updates, 2) positions of sparse updates. Reference \cite{sparse} presents a scheme that achieves information theoretic privacy of the values and positions of the sparse updates in the context of federated submodel learning (FSL) \cite{pruw_jpurnal,ourICC,pruw,rd,dropout}. The scheme in \cite{sparse} incurs a significant storage cost. In this work, we extend the scheme in \cite{sparse} to FL, with an additional variable that allows the storage cost to decrease at the expense of a certain amount of privacy leakage.

\begin{figure}[t]
    \centering
    \includegraphics[scale=0.5]{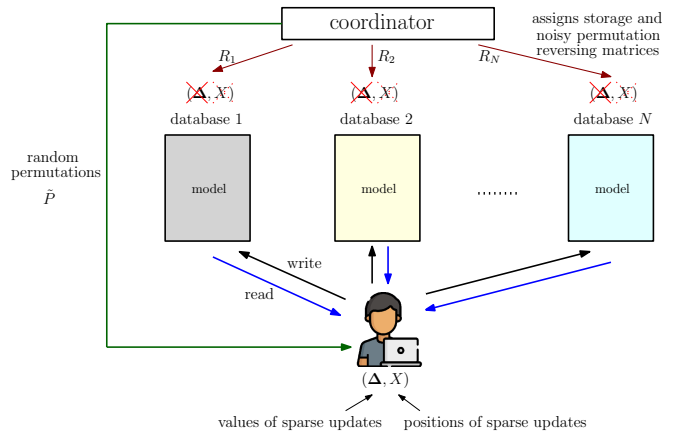}
    \caption{{System model.}}
    \label{model}
    \vspace*{-0.5cm}
\end{figure}

In this paper, we consider an FL setting with multiple non-colluding databases storing the ML model, and a user that communicates with the databases as shown in Fig.~\ref{model}. The schemes we propose are based on permutation techniques, where a coordinator initializes a random permutation of sets of parameters, and sends it to the users. The coordinator then places \emph{noise added permutation reversing matrices} at each database in such a way that the databases learn nothing about the underlying permutation. All communications between the user and the databases take place in terms of the permuted indices, which guarantee the privacy of the positions of the sparse updates. Despite the added noise which ensures privacy, the parameters in each database get placed in the correct place. 

\begin{figure}[t]
    \centering
    \includegraphics[scale=0.45]{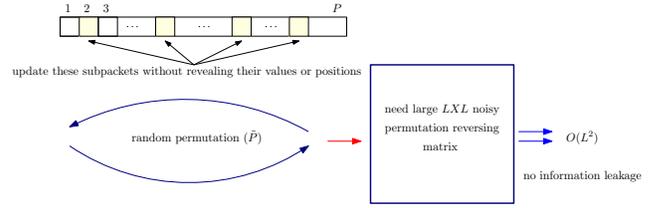}
    \caption{Without segmentation, the process requires large noise added permutation reversing matrices, which increase the storage cost while guaranteeing zero privacy leakage.}
    \label{motivation}
\end{figure}

\begin{figure}[t]
    \centering
    \includegraphics[scale=0.5]{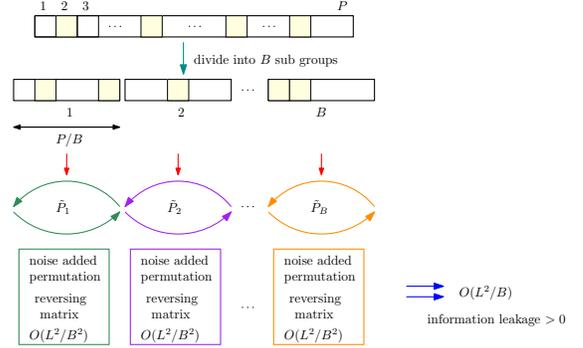}
    \caption{With segmentation, the process requires small noisy permutation reversing matrices resulting in a lower storage cost at the expense of a positive information leakage.}
    \label{idea}
    \vspace*{-0.5cm}
\end{figure}

The main challenge of this method is the significant storage cost incurred by the large permutation reversing matrices. We propose schemes that reduce the storage cost by reducing the size of the permutation reversing matrices, at the expense of a given amount of information leakage. This is achieved by dividing the ML model into multiple segments and carrying out permutations within each segment; see Figs.~\ref{motivation} and~\ref{idea}. The number of segments is chosen based on the allowed amount of information leakage and the storage capacity of the databases.

In this work, we propose two schemes to perform private FL with top $r$ sparsification for uncoded storage. We present the trade-off between the communication cost, storage complexity and information leakage in private FL with sparsification.

\section{Problem Formulation}\label{problem}
We consider an FL setting in which an ML model consisting of $L$ parameters belonging to $P$ subpackets is stored in $N$ non-colluding databases. The parameters take values from a large enough finite field $\mathbb{F}_q$. A given user at a given time $t$ trains the model using the user's local data. We consider sparsification in both uplink and  downlink, to reduce the communication cost. In particular, the sparsification rates of the reading (downlink) and writing (uplink) phases are given by $r'$ and $r$, respectively. In the reading (download) phase, the users only download a selected set of $Pr'$ subpackets determined by the databases.\footnote{These subpackets are determined based on the sparse updates received at time $t-1$, or by any other downlink sparsification protocol.} In the writing (upload) phase, each user only uploads the most significant $Pr$ subpackets to the databases.\footnote{We assume that all parameters in the sparse set of $Pr$ subpackets in the writing phase have non-zero updates.}

Note that  the users send no information to the databases in the reading phase. Therefore, no information about the user's local data is leaked to the databases in the reading phase. However, the users send the sparse updates and their positions (indices) to the databases in the writing phase to train the model. Information about the user's local data can be leaked to the databases from these updates and their indices.
In this work, we consider the following privacy guarantees on the values and the positions of the sparse updates.

\emph{Privacy of the values of sparse updates:} No information on the values of the sparse updates is allowed to leak to any of the databases, i.e., 
\begin{align}
    I(\Delta_i^{[t]};G_n^{[t]})=0, \quad n\in\{1,\dotsc,N\},
\end{align}
where $\Delta_{i}^{[t]}$ is the $i$th sparse (non-zero) update of a given user at time $t$ and $G_{n}^{[t]}$ contains all the information sent by the user to database $n$ at time $t$.

\emph{Privacy of the positions (indices) of sparse updates:} The amount of information leaked about the positions of the sparse updates need to be maintained under a given privacy leakage budget, i.e.,
\begin{align}
    I(X^{[t]};G_n^{[t]})<\epsilon, \quad n\in\{1,\dotsc,N\}, 
\end{align}
where $X^{[t]}$ is the set of indices of the sparse subpackets updated by a given user at time $t$. The system model with the privacy constraints is shown in Fig.~\ref{model}. A coordinator is used to initialize the scheme. In addition to the privacy constraints, we require the following security and correctness conditions for the reliability of the scheme.

\emph{Security of the model:} No information about the model parameters is allowed to leak to the databases, i.e.,
\begin{align}
    I(W^{[t]};S_n^{[t]})=0, \quad n\in\{1,\dotsc,N\},
\end{align}
where $W^{[t]}$ is the ML model and $S_n^{[t]}$ is the data content in database $n$ at time $t$.

\emph{Correctness in the reading phase:} The user should be able to correctly decode the sparse set of subpackets (denoted by $J$) of the model, determined by the downlink sparsification protocol, from the downloads in the reading phase, i.e., 
\begin{align}
H(W_{J}^{[t-1]}|A_{1:N}^{[t]})=0,
\end{align}
where $W_{J}^{[t-1]}$ is the set of subpackets in  set $J$ of the model $W$ at time $t-1$ (before updating) and $A_n^{[t]}$ is the information downloaded from database $n$ at time $t$.

\emph{Correctness in the writing phase:} Let $J'$ be the set of most significant $Pr$ subpackets of the model, updated by a given user at time $t$. Then, the model must be correctly updated as,
\begin{align}
    W_{s}^{[t]}=
    \begin{cases}
    W_{s}^{[t-1]}+\Delta_{s}^{[t]}, & \text{if $s\in J'$}\\
    W_{s}^{[t-1]}, & \text{if $s\notin J'$}
    \end{cases},
\end{align}
where $W_{s}^{[t-1]}$ is the subpacket $s$ of the ML model at time $t-1$ and $\Delta_{s}^{[t]}$ is the update of subpacket $s$ at time $t$.

\emph{Reading and writing costs:} The reading and writing costs are defined as $C_R=\frac{\mathcal{D}}{L}$ and $C_W=\frac{\mathcal{U}}{L}$, respectively, where $\mathcal{D}$ is the total number of symbols downloaded in the reading phase, $\mathcal{U}$ is the total number of symbols uploaded in the writing phase, and $L$ is the size of the model. The total cost $C_T$ is the sum of the reading and writing costs, $C_T=C_R+C_W$.

\emph{Storage complexity:} The storage complexity is quantified by the order of the total number of symbols stored in each database. 

In this work, we propose schemes to perform FL with top $r$ sparsification, that result in the minimum total communication cost and storage complexity, while satisfying all privacy, security and correctness conditions described above.

\begin{table*}[ht]
\begin{center}
\caption{Achievable sets of communication costs, storage costs, and amounts of information leakage.}
\label{main_res}
\vspace*{-0.1cm}
\begin{tabular}{ |c|c|c|c|c| }
\hline
  case & reading cost & writing cost & storage complexity & information leakage\\ 
  \hline
  Case 1 & $\frac{2r'(1+\frac{\log_q P}{N})}{1-\frac{2}{N}}$ & $\frac{2r(1+\log_q P)}{1-\frac{2}{N}}$ & $O(\frac{L^2}{B})$ & $H(\hat{X}_1,\dotsc,\hat{X}_B)$\\ 
  \hline
  Case 2 & $\frac{2r'(1+\frac{\log_q P}{N})}{1-\frac{4}{N}}$ & $\frac{2r(1+\log_q P)}{1-\frac{4}{N}}$ & $\max\{O(\frac{L^2}{B}),O(\frac{L^2B^2}{N^2})\}$ & $H(\Tilde{X}_1,\dotsc,\Tilde{X}_B)$\\
  \hline
\end{tabular}
\end{center}
\vspace*{-0.5cm}
\end{table*}

\section{Main Result}\label{main}
\begin{theorem}\label{main_result}
Consider an FL model stored in $N$ non-colluding databases, consisting of $L$ symbols from a finite field $\mathbb{F}_q$, which are included in $P$ subpacekts. The model is divided into $B$ segments of equal size ($1\leq B<P$), such that each consecutive $\frac{P}{B}$ subpackets are included in each segment. Assume that the FL model is updated by users at each time instance with uplink and downlink sparsification rates (top $r$ sparsification) of $r$ and $r'$, respectively. Let $\hat{X}_i$ be the random variable representing the number of subpackets with non-zero (sparse) updates of the $i$th segment generated by any given user, and let $(\Tilde{X}_1,\dotsc,\Tilde{X}_B)$ be the random vector representing all distinct combinations of $(\hat{X}_1,\dotsc,\hat{X}_B)$, irrespective of the segment index. Then, the reading/writing costs, storage complexities and amounts of information leakage presented in Table~\ref{main_res} are achievable.
\end{theorem}

\begin{remark}
When $B=1$ (no segmentation), $\hat{X}_1=\Tilde{X}_1=Pr$ and the corresponding infromation leakage is zero since $Pr$ is fixed and $H(\hat{X}_1)=H(\Tilde{X}_1)=0$. That is, the schemes corresponding to the two cases achieve information theoretic privacy of the values and positions of the sparse updates while incurring the same communication costs stated in Table~\ref{main_res}. 
\end{remark}

\begin{remark}
For a given privacy budget on the positions of the sparse updates given by $\epsilon$, the optimum number of segments $B$ can be calculated by minimizing the storage complexity, such that $H(\hat{X}_1,\dotsc,\hat{X}_B)<\epsilon$ or $H(\tilde{X}_1,\dotsc,\Tilde{X}_B)<\epsilon$ is satisfied. 
\end{remark}

\begin{remark}    
Since $H(\hat{X}_1, \dotsc, \hat{X}_B)$ considers all possible values of $\hat{X}_i$, while $H(\Tilde{X}_1,\dotsc,\Tilde{X}_B)$ only considers distinct sets of $\{\hat{X}_i\}_{i=1}^B$, $H(\hat{X}_1,\dotsc,\hat{X}_B)>H(\Tilde{X}_1,\dotsc,\Tilde{X}_B)$. 
\end{remark}

\begin{figure}[t]
    \centering
    \includegraphics[scale=0.45]{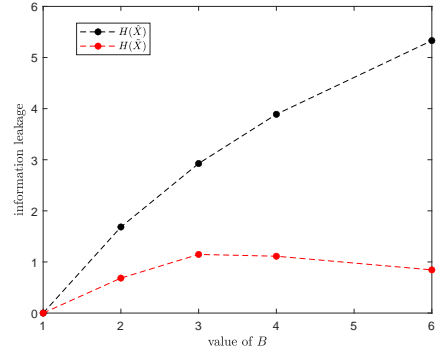}
    \caption{Information leakage of an example setting with $P=12$ versus $B$.}
    \label{leak}
    \vspace*{-0.5cm}
\end{figure}

\begin{remark}
Consider an example setting with $P=12$ subpackets divided into $B=1, 2, 3, 4, 6$ segments. Assume that each subpacket is equally probable to be selected to the set of most significant $Pr=3$ subpackets. The behavior of the information leakage for each value of $B$ is shown in Fig.~\ref{leak}.
\end{remark}

\begin{remark}
For the first case, one can achieve a lower storage cost at the expense of a higher information leakage by increasing $B$ and vice versa. This is because the number of different realizations of the placements of the $Pr$ sparse subpackets at the $B$ segments increases with $B$ when all permutations of the placements are considered. However, when only the distinct placements are considered in case 2 (without permutations), after $B=Pr$, the probability of each realization increases, which in turn decreases the entropy $H(\tilde{X}_1,\dotsc,\tilde{X}_B)$, i.e., the information leakage. Therefore, the storage-privacy leakage trade-off in case 2 also follows an inverse relation. The variation of the privacy leakage and the storage cost is controlled by the parameter $B$. The communication cost however is independent of $B$, which makes it independent of the storage cost and the privacy leakage.
\end{remark}

\section{Proposed Scheme}\label{proposedscheme}
We propose two schemes. Both schemes achieve privacy of the indices of the sparse subpackets by utilizing a permutation technique. In case 1, the model is divided into $B$ segments, and the scheme only considers permutations among the subpackets within each segment while revealing the real segment indices of the sparse subpackets. The scheme in case 2 considers permutations within and among segments to reduce the information leakage further. The schemes are presented in terms of examples due to space limitations here. 

\subsection{Case 1: Single Stage Permutations}
Consider an example setting with $P=15$ and $B=3$.

\subsubsection{Initialization}
\begin{figure}
		\centering
		\includegraphics[scale=0.55]{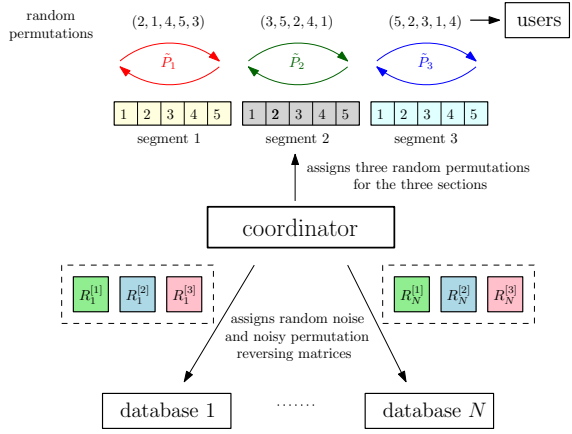}
		\caption{Initialization of the scheme for case 1.}
		\label{init_c1_eg}
		\vspace*{-0.5cm}
\end{figure}

The storage of a single subpacket $s$ in case 1 is given by,
\begin{align}\label{storage_c1_eg}
 S_n=\begin{bmatrix}
    \frac{1}{f_{1}-\alpha_n}W_{1}^{[s]}+\sum_{j=0}^{x}\alpha_n^j I_{1,j}\\
    \vdots\\
    \frac{1}{f_{\ell}-\alpha_n}W_{\ell}^{[s]}+\sum_{j=0}^{x}\alpha_n^j I_{\ell,j}
    \end{bmatrix},
\end{align}
where $x=\ell$,\footnote{$\ell$ is the subpacketization, for which an expression is derived at the end of Section~\ref{c1read}.} $W_i^{[s]}$ is the $i$th symbol of subpacket $s$, $I_{i,j}$ are random noise symbols and $\{f_i\}_{i=1}^{\ell},\{\alpha_n\}_{n=1}^N$ are globally known distinct constants from $\mathbb{F}_q$. At the initialization stage the coordinator sends $B=3$ randomly and independently chosen permutations of the $\frac{P}{B}=5$ subpackets in each of the $B=3$ segments to all users, denoted by $\tilde{P}_1,\tilde{P}_2,\tilde{P}_3$ as shown in Fig.~\ref{init_c1_eg}. The coordinator also sends the  corresponding noise added permutation reversing matrices given by,
\begin{align}
    R^{[i]}_n=\tilde{R}_n^{[i]}+\bar{Z}_i, \quad i=1,\dotsc,B,
\end{align}
to database $n$, $n\in\{1,\dotsc,N\}$, as shown in Fig.~\ref{init_c1_eg}, where $\tilde{R}_n^{[i]}$ is the scaled permutation reversing matrix corresponding to the permutation $\tilde{P}_i$ and $\tilde{Z}_i$ is a random noise matrix of size $\frac{P\ell}{B}\times\frac{P\ell}{B}$. Based on this example, the permutation reversing matrix for database $n$, $n\in\{1,\dotsc,N\}$ corresponding to the first segment (permutation: $\Tilde{P}_1=(2,1,4,5,3)$) is given by,
\begin{align}\label{R_1}
    R_n^{[1]}=\begin{bmatrix}
        0_{\ell\times\ell} & \Gamma_n & 0_{\ell\times\ell} & 0_{\ell\times\ell} & 0_{\ell\times\ell}\\
        \Gamma_n & 0_{\ell\times\ell} & 0_{\ell\times\ell} & 0_{\ell\times\ell} & 0_{\ell\times\ell}\\
        0_{\ell\times\ell} & 0_{\ell\times\ell} & 0_{\ell\times\ell}  & 0_{\ell\times\ell}& \Gamma_n\\
        0_{\ell\times\ell} & 0_{\ell\times\ell} & \Gamma_n & 0_{\ell\times\ell}&  0_{\ell\times\ell} \\
        0_{\ell\times\ell}  & 0_{\ell\times\ell} & 0_{\ell\times\ell} & \Gamma_n & 0_{\ell\times\ell}\\
    \end{bmatrix}+\bar{Z}_1,
\end{align}
where $\Gamma_n=\diag\{\frac{1}{f_1-\alpha_n}, \dots, \frac{1}{f_\ell-\alpha_n} \}$ and $0_{\ell\times\ell}$ is the all zeros matrix of size $\ell\times\ell$. 

\subsubsection{Reading Phase}\label{c1read}
The databases decide the permuted indices of the $Pr'$ sparse subpackets to be sent to the users at time $t$ in the reading phase, by selecting the permuted indices of the most commonly updated $Pr'$ subpackets by all users at time $t-1$. Note that the databases are unaware of the real indices of the sparse subpackets updated by the users in the writing phase at each time instance and only work with the permuted indices in both phases. For example, let the sparse set of permuted subpacket indices chosen by the databases to be sent to the users corresponding to the first segment be $\Tilde{V}_1=\{1,3\}$. One designated database sends these permuted indices of each segment to the users. The users then find the real indices, using the known permutations, i.e., for segment 1, the real set of indices is given by $V_1=\{2,4\}$.

In order to send the $i$th sparse subpacket of segment $j$ (permuted) denoted by $\Tilde{V}_j(i)$, each database $n$, $n\in\{1,\dotsc,N\}$ generates the following queries.
\begin{align}
    Q_n^{[\Tilde{V}_j(i)]}=\sum_{k=1}^\ell R_n^{[j]}(:, (i-1)\ell+k).
\end{align}
For example, the query corresponding to the first sparse subpacket of the first segment is given by,
\begin{align}
    Q_n^{[\Tilde{V}_1(1)]}=Q_n^{[1]}&=\sum_{k=1}^\ell R_n^{[1]}(:,k)=\begin{bmatrix}
        0_{\ell}\\\frac{1}{f_1-\alpha_n}\\\vdots\\\frac{1}{f_\ell-\alpha_n}\\0_{\ell}\\0_{\ell}\\0_{\ell}\\
    \end{bmatrix}+Z_1,
\end{align}
where $Z_1$ is a random noise vector resulted by the noise component of $R_n^{[1]}$. Then, database $n\in\{1,\dotsc,N\}$ sends the corresponding answer by calculating the dot product between the query and the scaled storage of the respective segment as,
\begin{align}
     A_n^{[\Tilde{V}_1(1)]}&\!=\!(D_n\times S_n)^T Q_n^{[\Tilde{V}_1(1)]}\\
     &\!=\!\frac{1}{f_1-\alpha_n}W_1^{[2]}+\dotsc+\frac{1}{f_\ell-\alpha_n}W_\ell^{[2]}+P_{\alpha_n}(\ell+1)\label{general_r},
\end{align}
where $D_n$ is the diagonal matrix of size $\frac{P\ell}{B}\times\frac{P\ell}{B}$ given by $D_n=\diag\{\Gamma_n^{-1},\dots,\Gamma_n^{-1}\}$ and $P_{\alpha_n}(\ell+1)$ is a polynomial in $\alpha_n$ of degree $\ell+1$.
Then, the users obtain the parameters of real subpacket 2, by solving the $N$ equations (answers from $N$ databases) of the form \eqref{general_r}, given that $N=2\ell+2$, which determines the subpacketization as $\ell=\frac{N-2}{2}$. The same procedure is carried out for all sparse subpackets in all segments. The resulting reading cost is given by,
\begin{align}\label{readcost}
    C_R\!=\!\frac{Pr'(\log_q \frac{P}{B}\!+\!\log_q B)+Pr'N}{L}\!=\!\frac{2r'(1\!+\!\frac{\log_q P}{N})}{1-\frac{2}{N}}.
\end{align}

\subsubsection{Writing Phase}
In the writing phase, each user generates non-zero updates for $Pr$ most significant subpackets, and sends the noise added combined updates (single symbol per subpacket) along with their permuted subpacket indices and the segment indices to each of the databases. The combined update of some (real) subpacket $i$ of segment $j$ (to be sent to database $n$, $n\in\{1,\dotsc,N\}$) is given by,
\begin{align}\label{upd}
    U_n^{[i,j]}\!=\!\!\sum_{k=1}^\ell\prod_{r=1,r\neq k}^\ell \!\!\!(f_r-\alpha_n)\Tilde{\Delta}_{k}^{[i,j]}\!+\!\prod_{r=1}^\ell(f_r-\alpha_n)Z^{[i,j]},
\end{align}
where $\Tilde{\Delta}_{k}^{[i,j]}=\frac{\Delta_{k}^{[i,j]}}{\prod_{r=1,r\neq k}^\ell(f_r-f_k)}$ with $\Delta_{k}^{[i,j]}$ being the update of the $k$th symbol of subpacket $i$ of segment $j$ and $Z^{[i,j]}$ is a random noise symbol. Note that the addition of $Z^{[i,j]}$ to the updates in \eqref{upd} guarantees information theoretic privacy of the values of updates from Shannon's one time pad theorem. For example, assume that a given user wants to update the real subpackets 2 and 4 from segment 1, subpacket 2 from segment 2, and subpacket 5 from segment 3. Based on the permutations considered in this example, i.e., $\Tilde{P}_1=\{2,1,4,5,3\}$, $\Tilde{P}_2=\{3,5,2,4,1\}$ and $\Tilde{P}_3=\{5,2,3,1,4\}$, the user generates the permuted (update, subpacket, segment) tuples given by $(U_n^{[2,1]},1,1)$, $(U_n^{[4,1]},3,1)$ for segment 1, $(U_n^{[2,2]},3,2)$ for segment 2, and $(U_n^{[5,3]},1,3)$ for segment 3. Note that there is no permutation in the segment index, and only the subpacket indices within each segment are being permuted. Database $n$, $n\in\{1,\dotsc,N\}$ creates permuted update vectors for each segment upon receiving the $Pr$ permuted (update, subpacket, segment) tuples. For segment 1, the permuted update vector is given by, 
\begin{align}
    \hat{U}_n^{[1]}\!=\![U_n^{[2,1]}\cdot 1_\ell^T, \ 0\cdot 1_\ell^T, \ U_n^{[4,1]}\cdot 1_\ell^T, \ 0\cdot 1_\ell^T, \ 0\cdot 1_\ell^T]^T
\end{align}
where $1_\ell$ is the all ones vector of size $\ell\times1$. Next, the databases privately rearrange the updates in the real order and calculate the incremental updates of each segment. The incremental update calculation of segment 1 in database $n$ is given by,
\begin{align}
    \bar{U}_n^{[1]}&\!=\!R_n^{[1]}\hat{U}_n^{[1]}\\
    &\!=\!\left(\!\!\begin{bmatrix}
        0_{\ell\times\ell} & \Gamma_n & 0_{\ell\times\ell} & 0_{\ell\times\ell} & 0_{\ell\times\ell}\\
        \Gamma_n & 0_{\ell\times\ell} & 0_{\ell\times\ell} & 0_{\ell\times\ell} & 0_{\ell\times\ell}\\
        0_{\ell\times\ell} & 0_{\ell\times\ell} & 0_{\ell\times\ell}  & 0_{\ell\times\ell}& \Gamma_n\\
        0_{\ell\times\ell} & 0_{\ell\times\ell} & \Gamma_n & 0_{\ell\times\ell}&  0_{\ell\times\ell} \\
        0_{\ell\times\ell}  & 0_{\ell\times\ell} & 0_{\ell\times\ell} & \Gamma_n & 0_{\ell\times\ell}\\
    \end{bmatrix}\!\!+\!\!\bar{Z}_1\!\!\right)\begin{bmatrix}\!\!        U_n^{[2,1]}\!\cdot\!1_\ell\\0\cdot 1_\ell\\U_n^{[4,1]}\!\cdot\!1_\ell\\0\cdot1_\ell\\0\cdot1_\ell
    \end{bmatrix}\\
    &\!=\!\begin{bmatrix}
        0_\ell\\\frac{U_n^{[2,1]}}{f_1-\alpha_n}\\\vdots\\\frac{U_n^{[2,1]}}{f_\ell-\alpha_n}\\0_\ell\\\frac{U_n^{[4,1]}}{f_1-\alpha_n}\\\vdots\\\frac{U_n^{[4,1]}}{f_\ell-\alpha_n}\\0_\ell
    \end{bmatrix}+P_{\alpha_n}(\ell)=\begin{bmatrix}
        0_\ell\\\frac{\Delta_{1}^{[2,1]}}{f_1-\alpha_n}\\\vdots\\\frac{\Delta_{\ell}^{[2,1]}}{f_\ell-\alpha_n}\\0_\ell\\\frac{\Delta_{1}^{[4,1]}}{f_1-\alpha_n}\\\vdots\\\frac{\Delta_{\ell}^{[4,1]}}{f_\ell-\alpha_n}\\0_\ell
    \end{bmatrix}+P_{\alpha_n}(\ell)\label{incr_c1},
\end{align}
where $P_{\alpha_n}(\ell)$ here is a vector of size $\frac{L}{B}$ consisting of polynomial in $\alpha_n$ of degree $\ell$, and the last equality is obtained by applying \cite[Lemma~1]{pruw_jpurnal}. The same process is carried out for the other two segments as well. Since the incremental update is in the same form as the storage in \eqref{storage_c1_eg}, the storage of segment $j$, $j\in\{1,2,3\}$ at time $t$ can be updated as,
\begin{align}
    S_n^{[j]}(t)=S_n^{[j]}(t-1)+\bar{U}_n^{[j]}, \quad n\in\{1,\dots,N\}.
\end{align}
Note from \eqref{incr_c1} that for segment 1, the two real sparse subpackets 2 and 4 have been correctly updated, while ensuring that the rest of the subpackets remain the same, without revealing the real subpacket indices 2 and 4 to any of the databases. The resulting writing cost is given by,
\begin{align}
    C_W\!=\!\frac{PrN(1+\log_q B+\log_q \frac{P}{B})}{L}\!=\!\frac{2r(1+\log_qP)}{1-\frac{2}{N}}.
\end{align}
The total storage complexity (data and permutation reversing matrices) is given by $O(L)+O(\frac{L^2}{B^2}\times B)=O(\frac{L^2}{B})$. 

\subsection{Case 2: Two-Stage Permutations}
In case 1, only the subpacket indices within each segment were permuted, and the real segment indices were uploaded to the databases by the users. In this case, we permute subpacket indices within segments as well as the segment indices to reduce the information leakage. However, this increases the storage cost since the permutation of segment indices requires an additional noise added permutation reversing matrix to be stored in the databases. For case 2, consider an example setting with $P=12$ subpackets (with subpacketization $\ell$) which are divided into and $B=3$ equal segments.

\begin{figure}
		\centering
		\includegraphics[scale=0.55]{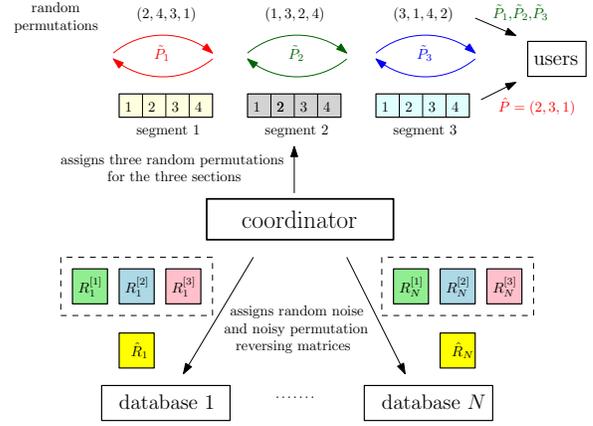}
		\caption{Initialization of the scheme for case 2.}
		\label{init_c3_eg}
		\vspace*{-0.5cm}
\end{figure}

\subsubsection{Initialization}
The storage of a single subpacket in this case is the same as \eqref{storage_c1_eg} with $x=\ell+1$. The coordinator places the $B=3$ permutations and the corresponding noisy permutation reversing matrices similar to case 1. In addition, the coordinator randomly and independently selects a permutation of the $B=3$ segments $\hat{P}$ and sends it to the users, while placing the corresponding noise-added permutation reversing matrix $\hat{R}_n$ at database $n$, $n\in\{1,\dotsc,N\}$. Consider the example setting given in Fig.~\ref{init_c3_eg}. The noise added permutation reversing matrix corresponding to the first segment ($\tilde{P}_1=\{2,4,3,1\}$) is given by,
\begin{align}\label{R1}
    R_n^{[1]}=\begin{bmatrix}
        0_{\ell\times\ell} & 0_{\ell\times\ell} & 0_{\ell\times\ell} & \Gamma_n\\
        \Gamma_n & 0_{\ell\times\ell} & 0_{\ell\times\ell} & 0_{\ell\times\ell} \\
        0_{\ell\times\ell} & 0_{\ell\times\ell} & \Gamma_n &0_{\ell\times\ell}\\
        0_{\ell\times\ell} & \Gamma_n & 0_{\ell\times\ell}&  0_{\ell\times\ell} 
    \end{bmatrix}+\bar{Z}_1,
\end{align}
with the same notation as in case 1. The noise added permutation reversing matrix at database $n$ corresponding to the segmentwise permutation ($\hat{P}=\{2,3,1\}$) is given by,
\begin{align}
\hat{R}_n&=\begin{bmatrix}
        0_{\ell\times\ell} & 0_{\ell\times\ell} & \Phi\\
        \Phi & 0_{\ell\times\ell} & 0_{\ell\times\ell}\\
        0_{\ell\times\ell} & \Phi & 0_{\ell\times\ell}\\ 
    \end{bmatrix}+\begin{bmatrix}
        \Gamma_n^{-1} & 0_{\ell\times\ell} & 0_{\ell\times\ell}\\
        0_{\ell\times\ell} & \Gamma_n^{-1} & 0_{\ell\times\ell}\\
        0_{\ell\times\ell} & 0_{\ell\times\ell} & \Gamma_n^{-1}\\ 
    \end{bmatrix}\hat{Z}\\
    &=\begin{bmatrix}
        b_{1,1}^{[n]} & b_{1,2}^{[n]} & b_{1,3}^{[n]}\\
        b_{2,1}^{[n]} & b_{2,2}^{[n]} & b_{2,3}^{[n]}\\
        b_{3,1}^{[n]} & b_{3,2}^{[n]} & b_{3,3}^{[n]}
    \end{bmatrix},\label{blocks}
\end{align}
where $\Phi=\text{diag}(1_\ell)$ and $\hat{Z}$ is a random noise matrix of size $B\ell\times B\ell$. Each matrix $\hat{R}_n$ is represented in blocks of size $\ell\times\ell$, as shown in \eqref{blocks}.

\subsubsection{Reading Phase}
In the reading phase, the databases determine the permuted indices of the $Pr'$ most significant subpackets to be sent to the users as described in case 1. Assume that the permuted (subpacket, segment) pair of a chosen subpacket is $(\eta_p,\phi_p)=(1,3)$. A designated database sends this information to the user and the user finds the corresponding real segment $\phi_r$ and subpacket $\eta_r$ as $\phi_r=\hat{P}(\phi_p)=1$ and $\eta_r=\tilde{P}_{\phi_r}(\eta_p)=2$. To send the corresponding real subpacket, the databases first generate the combined noisy permutation reversing matrix given by,
\begin{align}\label{comb}
    R_n&\!\!=\!\!\begin{bmatrix}
        R_n^{[1]} \!\!\!\!&\!\!\!\!  & \\
         &\!\!\!\! R_n^{[2]}\!\!\!\! & \\
         &\!\!\!\!  &\!\!\!\! R_n^{[3]}
    \end{bmatrix}\!\!\times\!\! \begin{bmatrix}
        \begin{bmatrix}
            b^{[n]}_{1,1}\!\!\!\! &\!\!\!\!   & \\
            &\!\!\!\!\ddots\!\!\!\!&\\
            & \!\!\!\! & \!\!\!\!b^{[n]}_{1,1}
        \end{bmatrix}
        \begin{bmatrix}
            b^{[n]}_{1,2}\!\!\!\! &\!\!\!\!   & \\
            &\!\!\!\!\ddots\!\!\!\! &\\
            & \!\!\!\! &\!\!\!\! b^{[n]}_{1,2}
        \end{bmatrix}
        \begin{bmatrix}
            b^{[n]}_{1,3}\!\!\!\! &\!\!\!\!   & \\
            &\!\!\!\!\ddots\!\!\!\! &\\
            & \!\!\!\! &\!\!\!\! b^{[n]}_{1,3}
        \end{bmatrix}\\
        \begin{bmatrix}
            b^{[n]}_{2,1}\!\!\!\! &\!\!\!\!   & \\
            &\!\!\!\!\ddots\!\!\!\! &\\
            & \!\!\!\! &\!\!\!\! b^{[n]}_{2,1}
        \end{bmatrix}
        \begin{bmatrix}
            b^{[n]}_{2,2}\!\!\!\! & \!\!\!\!  & \\
            &\!\!\!\!\ddots\!\!\!\! &\\
            & \!\!\!\! & \!\!\!\!b^{[n]}_{2,2}
        \end{bmatrix}
        \begin{bmatrix}
            b^{[n]}_{2,3}\!\!\!\! &\!\!\!\!   & \\
            &\!\!\!\!\ddots\!\!\!\! &\\
            & \!\!\!\! &\!\!\!\! b^{[n]}_{2,3}
        \end{bmatrix}\\
        \begin{bmatrix}
            b^{[n]}_{3,1}\!\!\!\! &\!\!\!\!   & \\
            &\!\!\!\!\ddots\!\!\!\! &\\
            & \!\!\!\! & \!\!\!\!b^{[n]}_{3,1}
        \end{bmatrix}
        \begin{bmatrix}
            b^{[n]}_{3,2} \!\!\!\!& \!\!\!\!  & \\
            &\!\!\!\!\ddots\!\!\!\! &\\
            & \!\!\!\! &\!\!\!\! b^{[n]}_{3,2}
        \end{bmatrix}
        \begin{bmatrix}
            b^{[n]}_{3,3} \!\!\!\!& \!\!\!\!  & \\
            &\!\!\!\!\ddots\!\!\!\! &\\
            & \!\!\!\! &\!\!\!\! b^{[n]}_{3,3}
        \end{bmatrix}
    \end{bmatrix}.
\end{align}
In order to send the permuted subpacket indicated by $(\eta_p,\phi_p)=(i,j)$, each database generates the query given by,
\begin{align}
Q_n^{[i,j]}\!=\!\begin{bmatrix}\!
        \Gamma_n^{-1}\!\!\!\! &\!\!\!\! &\\
        &\!\!\!\! \ddots\!\!\!\! &\\
        &\!\!\!\! & \!\!\!\!\Gamma_n^{-1}\!\!\!
    \end{bmatrix}_{L\times L}\!\!\!\!\!\times \sum_{k=1}^\ell \! R_n(:,(j\!-\!1)\frac{P\ell}{B}\!+\!(i\!-\!1)\ell\!+\!k).
\end{align}
Then, the answer is generated by the dot product between the query and the storage as explained in case 1. In order for the user to be able to download the required subpacket using the $N$ answers, the system should satisfy $N=2\ell+4$, fixing the subpacketization of case 2 at $\ell=\frac{N-4}{2}$, which results in the reading cost given in Table~\ref{main_res} for case 2, using a similar calculation as in \eqref{readcost}.

\subsubsection{Writing Phase}
In the writing phase, the user sends the combined updates, permuted subpacket indices and permuted segment indices of the $Pr$ most significant subpackets to all databases. Assume that a given user wants to update the $Pr$ sparse subpackets identified by the real (subpacket, segment) pairs given by, $(\eta_r,\phi_r)=\{(2,1),(1,2),(3,3)\}$. Based on within segment permutations given by $\Tilde{P}_1=(2,4,3,1)$, $\Tilde{P}_2=(1,3,2,4)$, $\Tilde{P}_3=(3,1,4,2)$, and the segmentwise permutation given by $\hat{P}=(2,3,1)$, the user sends the following (permuted) information to database $n$, $n\in\{1,\dots,N\}$,
\begin{align}\label{rec_c3}
    (U_n^{[\eta_r,\phi_r]},\eta_p,\phi_p)=\{(U_n^{[2,1]},1,3),(U_n^{[1,2]},1,1),(U_n^{[3,3]},1,2)\}
\end{align}
where the combined updates $U_n^{[\eta_r,\phi_r]}$ are of the form \eqref{upd}. Once the databases receive all permuted (update, subpacket, segment) tuples, they construct the permuted update vector as,
\begin{align}
    \tilde{U}_n=[U_n^{[1,2]},0,0,0,U_n^{[3,3]},0,0,0,U_n^{[2,1]},0,0,0]^T
\end{align}
This vector is then scaled by an all ones vector of size $\ell\times1$ to aid the rest of the calculations. The scaled permuted update vector is given by $\hat{U}_n=[\tilde{U}_n(1)\cdot 1_\ell^T, \dotsc, \tilde{U}_n(12)\cdot 1_\ell^T]^T$. Then, database $n$, $n\in\{1,\dotsc,N\}$ calculates the incremental update using the combined noisy permutation reversing matrix in \eqref{comb} as $\bar{U}_n=R_n\times\hat{U}_n$, which is of the same form as the storage in \eqref{storage_c1_eg} with $x=\ell+1$. Therefore, the storage at time $t$, $S_n^{[t]}$ can be updated as $S_n^{[t]}=S_n^{[t-1]}+\bar{U}_n$. 

The storage complexities of data, noise added intra and inter segment permutation reversing matrices are given by $O(L)$, $O(\frac{L^2}{B})$ and $O(\ell^2B^2)=O(\frac{L^2B^2}{N^2})$, respectively. Therefore, the storage complexity is $\max\{O(\frac{L^2}{B}),O(\frac{L^2B^2}{N^2})\}$.

In the two proposed schemes, there exists a positive information leakage when $B>1$, since the numbers of subpackets with non-zero updates in each segment is revealed to the databases (in permuted or non-permuted order). This information leakage is characterized in terms of entropy expressions as shown in Table~\ref{main_res}, for the two cases separately. The proofs are omitted in this paper due to space limitations.  

\bibliographystyle{unsrt}
\bibliography{references}

\end{document}